\documentstyle[aps,eqsecnum,preprint,prd,epsf]{revtex}
\begin{document}
\addtolength{\jot}{10pt}
\tighten

\draft
\preprint{\vbox{\hbox{BARI-TH/95-206 \hfill} 
                \hbox{DSF-T-95/42 \hfill}
                \hbox{hep-ph/9510403 \hfill}
                 }}
\vskip 1cm
\title{\bf QCD Sum Rule Analysis of the Decays \\
$B \to K \ell^+ \ell^-$ and $B \to K^* \ell^+ \ell^-$ \\}
\vskip 0.5cm
\author{P.~Colangelo $^{1}$, F.~De~Fazio $^{1,2}$,
P.~Santorelli $^{3}$ and E.~Scrimieri $^{1,2}$\\ }

\vskip 1.0cm

\address{$^{1}$ Istituto Nazionale di Fisica Nucleare, \\  
 Sezione di Bari, Italy  \\
$^{2}$ Dipartimento di Fisica dell'Universit\`a di Bari,   \\
via G.Amendola 173, 70126 Bari, Italy \\
$^{3}$ Dipartimento di Fisica dell'Universit\`a ``Federico II'' di Napoli \\
and Istituto Nazionale di Fisica Nucleare, Sezione di Napoli,\\ 
Mostra D'Oltremare, Pad 19-20, 80125 Napoli, Italy}


\maketitle
\vskip 0.5cm
\begin{abstract}
We use QCD sum rules to calculate the hadronic
matrix elements governing the rare  decays  $B \to K \ell^+ \ell^-$ and 
$B \to K^* \ell^+ \ell^-$ induced by the flavour changing neutral
current $b \to s$ transition. We also study relations among semileptonic
and rare $B \to K^{(*)}$ decay form factors. The analysis of the
invariant mass distribution of the lepton pair in $B \to K^{(*)} \ell^+
\ell^-$ and of the angular asymmetry in $B \to K^* \ell^+ \ell^-$
provides us with interesting tests of the Standard Model and its
extensions. 
\end{abstract}

\vspace{1truecm}


\clearpage

\section{Introduction}

Rare $B$-meson decays induced by the flavour changing neutral current $b
\to s$ transition represent important channels for testing the Standard
Model (SM) and for searching the effects of possible new interactions
\cite{ras}. As a matter of fact, these processes, that in SM  do not
occur in the Born approximation, are particularly sensitive to
perturbative QCD corrections and to possible higher mass scales and
interactions predicted in supersymmetric theories, two Higgs doublet and
topcolor models, left-right models, etc. Such interactions determine the
operators and their Wilson coefficients appearing in the low energy
$\Delta B=1$ effective Hamiltonian $H_W$ that governs the $b \to s$
transition. \par From the experimental point of view, the radiative $b
\to s \gamma$ decay has been observed and measured by CLEO II
Collaboration both in the inclusive $B \to X_s \gamma$ and exclusive $B
\to K^* \gamma$ modes; the experimental results 
\begin{equation}
B(b \to s \gamma)=(2.32 \pm 0.57 \pm 0.35) \; 10^{-4} \;\;\;\;\; \cite{CLEOa} 
\label{inclusivo}
\end{equation}
\noindent and
\begin{equation}
\begin{array}{c}
\displaystyle
B({\bar B}^0 \to K^{*0} \gamma)=(4.0 \pm 1.7 \pm 0.8) \; 10^{-5} \\
\\
\displaystyle
B(B^- \to K^{*-} \gamma)=(5.7 \pm 3.1 \pm 1.1) \; 10^{-5}  
\end{array}
\;\;\;\;\cite{CLEOb} 
\label{esclusivo} 
\end{equation}
\noindent have prompted a number of analyses aimed at restricting the parameter 
space of various extensions of the Standard Model \cite{hew}. Similar 
analyses have also been proposed for the transition 
$b \to s \ell^+ \ell^-$, that 
has not been observed, yet \cite{ali}; in this case, 
 the invariant dilepton mass distribution and the asymmetry of the 
dilepton angular distribution, together with the total decay rate, 
 can be used to study the features of the 
interaction inducing the decay. However, for the exclusive modes such as $B \to 
K \ell^+ \ell^-$ and  $B \to K^* \ell^+ \ell^-$  one has to face the problem of 
computing the matrix element of $H_W$ between the states $B$ and $K$, 
$K^*$, a problem related to the nonperturbative sector of QCD. \par
For these matrix elements, either specific hadronization models 
\cite{desh,wyler} or 
information from two point function QCD sum rules \cite{paver} 
and from the heavy meson 
chiral theory \cite{nar}, embedded in the vector meson dominance framework, 
have been used so far. 
The resulting theoretical predictions 
are characterized by a considerable 
model dependence; it should be noticed that, differently from the case of 
$B \to K^* \gamma$, where the hadronic matrix element must be computed 
only at one kinematical point, in correspondence to the on-shell photon, for  
$B \to K \ell^+ \ell^-$ and  $B \to K^* \ell^+ \ell^-$ the matrix elements must 
be known in a wide range of the invariant mass squared of the lepton pair:
$M^2_{\ell^+ \ell^-}=[4M_\ell^2, (M_B-M_{K,K^*})^2]$; 
therefore, the vector meson dominance assumption has not negligible 
consequences on 
the theoretical outcome. \par
An approach based on general features of QCD that allows us to compute the 
hadronic matrix elements in a range of $M^2_{\ell^+ \ell^-}$ 
is provided by three-point function
QCD sum rules \cite{shiflibro}. 
This method, first employed to compute the pion form factor 
\cite{ioffe},  has been widely applied to 
heavy meson semileptonic decays: for example, in the case of $B \to D,D^*$ 
semileptonic transitions, it has been used to compute the Isgur-Wise universal 
function $\xi(y)$ and the heavy quark mass corrections \cite{neub}. 
Moreover, the decays $B \to D^{**} \ell \nu$, where $D^{**}$ are positive 
parity $(c {\bar q})$ meson states,
 have been analyzed both for finite heavy quark masses \cite{ovch}
and in the limit $m_Q \to \infty$, with the calculation of the universal 
functions $\tau_{1\over 2}(y)$ and $\tau_{3\over 2}(y)$ 
analogous to the Isgur-Wise function \cite{tau}.
For the heavy-to-light meson 
transitions, such  as $D(B) \to \pi(\rho)\ell \nu$, the various matrix elements 
have also been computed \cite{dosch,ball}; in 
the case of $B \to K^* \gamma$, this approach,
employed in \cite{noi,ball1,naris}, has provided us with the 
prediction $R={{B}(B \to K^* \gamma) / { B}(b \to s \gamma)} 
= 0.17\pm 0.05 $ \cite{noi}, that
agrees with the central value obtained from 
the experimental data in eqs.(\ref{inclusivo})-(\ref{esclusivo}).\par
In this paper we want to apply the 
three-point function QCD sum rule method to compute the 
hadronic quantities appearing in the calculation of $B \to K^{(*)} \ell^+ 
\ell^-$. We shall observe that the various form factors 
parametrizing the relevant matrix elements have common
features with other heavy-to-light meson transitions, 
a behaviour whose origin is worth investigating in detail \cite{stech}. 
We shall also compare the computed
hadronic quantities to the findings  of 
lattice QCD, even though these last results are obtained 
after extrapolations in the heavy quark mass and in the momentum transfer. 
Finally, we shall apply our results to predict 
the invariant mass distribution of the lepton pair in the decays
$B \to K^{(*)} \ell^+ \ell^-$ and the forward-backward 
asymmetry for $B \to K^* \ell^+ \ell^-$ in the Standard Model. 

\par
The work is organized as follows: in Sec. II we write down the 
(SM) effective 
Hamiltonian for the transition $b \to s \ell^+ \ell^-$, and 
resume the available information on the Wilson coefficients. In Sec. III we 
compute by three-point function QCD 
sum rules the relevant hadronic quantities
for $B \to K \ell^+ \ell^-$; the same calculation is 
carried out for  $B \to K^* \ell^+ \ell^-$ in Sec. IV.
In Sect. V we study the relations derived by Isgur and Wise \cite{IW}
and Burdman and Donoghue \cite{BD} between rare and semileptonic
form factors. Such relations can be worked out in the infinite heavy quark 
mass limit $m_b \to \infty$, in the region of maximum momentum transfer $t$;
a relevant problem is whether they are satisfied 
also in the low $t$ region, as it has been argued by several authors. 
We investigate this hypothesis and comment on the role of the
heavy mass corrections.
In Sec. VI and VII we study the transitions
$B \to K \ell^+ \ell^-$ and $B \to K^* \ell^+ \ell^-$, respectively. 
Finally, in Sec. VIII we draw our conclusions. 
Details concerning the calculations are reported in the Appendix.

\section{Effective Hamiltonian}

The effective $ \Delta B =-1$, $\Delta S = 1$ 
Hamiltonian governing in the Standard Model the rare
transition $b \to s \ell^+ \ell^-$ can be written in terms of a 
set of local operators \cite{gri}: 
\begin{equation}
H_W\,=\,4\,{G_F \over \sqrt{2}} V_{tb} V_{ts}^* \sum_{i=1}^{10} C_i(\mu)
O_i(\mu) 
\label{hamil} 
\end{equation}
\noindent where $G_F$ is the Fermi constant and $V_{ij}$ are elements of the 
Cabibbo-Kobayashi-Maskawa mixing matrix; we neglect terms proportional 
to $V_{ub} V_{us}^*$ since the ratio $\displaystyle \left 
|{V_{ub}V_{us}^* \over V_{tb}V_{ts}^*}\right |$ 
is of the order $10^{-2}$. The operators $O_i$, written in terms 
of quark and gluon fields, read as follows:
\begin{eqnarray}
O_1&=&({\bar s}_{L \alpha} \gamma^\mu b_{L \alpha})
      ({\bar c}_{L \beta} \gamma_\mu c_{L \beta}) \nonumber \\
O_2&=&({\bar s}_{L \alpha} \gamma^\mu b_{L \beta})
      ({\bar c}_{L \beta} \gamma_\mu c_{L \alpha}) \nonumber \\
O_3&=&({\bar s}_{L \alpha} \gamma^\mu b_{L \alpha})
      [({\bar u}_{L \beta} \gamma_\mu u_{L \beta})+...+
      ({\bar b}_{L \beta} \gamma_\mu b_{L \beta})] \nonumber \\
O_4&=&({\bar s}_{L \alpha} \gamma^\mu b_{L \beta})
      [({\bar u}_{L \beta} \gamma_\mu u_{L \alpha})+...+
      ({\bar b}_{L \beta} \gamma_\mu b_{L \alpha})] \nonumber \\
O_5&=&({\bar s}_{L \alpha} \gamma^\mu b_{L \alpha})
      [({\bar u}_{R \beta} \gamma_\mu u_{R \beta})+...+
      ({\bar b}_{R \beta} \gamma_\mu b_{R \beta})] \nonumber \\
O_6&=&({\bar s}_{L \alpha} \gamma^\mu b_{L \beta})
      [({\bar u}_{R \beta} \gamma_\mu u_{R \alpha})+...+
      ({\bar b}_{R \beta} \gamma_\mu b_{R \alpha})] \nonumber \\
O_7&=&{e \over 16 \pi^2} m_b ({\bar s}_{L \alpha} \sigma^{\mu \nu} 
     b_{R \alpha}) F_{\mu \nu} \nonumber \\
O_8&=&{g_s \over 16 \pi^2} m_b \Big[{\bar s}_{L \alpha} \sigma^{\mu \nu} 
      \Big({\lambda^a \over 2}\Big)_{\alpha \beta} b_{R \beta}\Big] \; 
      G^a_{\mu \nu} \nonumber\\
O_9&=&{e^2 \over 16 \pi^2}  ({\bar s}_{L \alpha} \gamma^\mu 
     b_{L \alpha}) \; {\bar \ell} \gamma_\mu \ell \nonumber \\
O_{10}&=&{e^2 \over 16 \pi^2}  ({\bar s}_{L \alpha} \gamma^\mu 
     b_{L \alpha}) \; {\bar \ell} \gamma_\mu \gamma_5 \ell 
\label{eff} 
\end{eqnarray}
\noindent ($\alpha$, $\beta$ are 
colour indices, $\displaystyle b_{R,L}={1 \pm \gamma_5 \over 2}b$, and
$\displaystyle \sigma^{\mu \nu}={i \over 2}[\gamma^\mu,\gamma^\nu]$); 
$e$ and $g_s$ are the electromagnetic and the strong coupling constant,
respectively,  $F_{\mu \nu}$ and $G^a_{\mu \nu}$ in $O_7$ and $O_8$
denote the electromagnetic and the gluonic field strength tensor. $O_1$
and $O_2$ are current-current operators, $O_3,...,O_6$ are usually named
QCD penguin operators, $O_7$ (inducing the radiative $b \to s \gamma$
decay) and $O_8$ are magnetic penguin operators, $O_9$ and $O_{10}$ are
semileptonic electroweak penguin operators. 

The Wilson coefficients $C_i(\mu)$ have been partially computed at the
next-to-leading order in QCD by several groups \cite{misiak,BurasMunz,qcd}.
As discussed in ref.\cite{BurasMunz}, in 
the analysis of $B\rightarrow X_s \ell^+\ell^-$ at 
the next-to-leading logarithmic corrections 
must be consistently included only in the coefficient $C_9$, since 
at the leading approximation
$O_9$ is the only operator responsible of the transition 
$b \rightarrow s~\ell^+ \ell^-$. The contribution of the other
operators (excluding $O_8$ that, however, 
is not involved in the processes we are studying)
appears only at the next-to-leading order, 
and therefore their Wilson coefficients must be evaluated at the leading 
approximation.\\
Following \cite{BurasMunz} we use in our phenomenological analysis 
of the decays $B \to K^{(*)} \ell^+ \ell^-$ (within the Standard Model) the 
numerical values of the Wilson coefficients collected in Table I. We choose
the scale $\mu=5 \; GeV \simeq m_b$, 
$\Lambda^{(5)}_{{\overline {MS}}}=225 \; MeV$ and  the top quark mass 
$m_t=174 \; GeV$ from the CDF measurement \cite{CDF}. The coefficient $C_9$, 
which is evaluated at the next-to-leading order approximation, displays a 
dependence on the regularization scheme, as it can be observed in Table I 
comparing the result obtained using the 't Hooft-Veltman (HV) and 
the Naive Dimensional Regularization (NDR) scheme.  Such dependence must 
disappear in the decay amplitude if all corrections are taken into account. We 
shall include in our analysis the uncertainty on $C_9$ as a part of the 
theoretical error. In Table I it can also be observed that the coefficients of 
$O_3-O_6$ are small (${\cal O}(10^{-2})$); therefore, the contribution of such 
operators can be neglected, and the analysis can be carried out considering 
only  the operators 
$O_1$, $O_2$ and to $O_7$, $O_9$ and $O_{10}$.\par
The various extensions of the Standard Model, such as models involving 
supersymmetry, multiHiggs and left-right models, induce two kind of changes
in the low energy Hamiltonian (\ref{hamil}): first, the values of the 
coefficients $C_i$ are modified as an effect of additional virtual 
particles in the loop diagrams describing the $b \to s$ transition, and, 
second, new operators can appear in the operator basis, 
such as operators with 
different chirality of the quark current with respect to $O_7-O_{10}$, e.g.,
$\displaystyle O^\prime_7={e \over 16 \pi^2} m_b ({\bar s}_R \sigma^{\mu
\nu} b_L) F_{\mu \nu}$, 
$\displaystyle O^\prime_9={e^2 \over 16 \pi^2}  ({\bar s}_R \gamma^\mu 
b_R){\bar \ell} \gamma_\mu \ell$ and 
$\displaystyle O^\prime_{10}={e^2 \over 16 \pi^2}  ({\bar s}_R
\gamma^\mu  b_R){\bar \ell} \gamma_\mu \gamma_5 \ell$. \par 
This rich structure justifies the interest for $B \to K^{(*)} \ell^+ \ell^-$, 
where operators of different origin  act coherently    in determining 
rates, spectra and asymmetries. For example, it could be interesting to search 
for 
the effects of possible interactions that produce a coefficient
$C_7$ with opposite sign \cite{ali,wyler}. In 
this work we shall not analyze such new effects, limiting ourselves to studying 
the above processes within the theoretical framework provided us by the 
Standard Model. However, it is worth stressing that our results for the 
hadronic matrix elements of the operators appearing in (\ref{hamil}) represent 
a complete set of quantities also for the analysis of the decays
$B \to K^{(*)} \ell^+ \ell^-$ in a context different from the Standard Model.

\section{Form factors of the decay $B \to K \ell^+ \ell^-$}

The matrix elements of the operators 
$O_1$, $O_2$ and $O_7$, $O_9$ and $O_{10}$
in eq.(\ref{eff}) between the external states $B$ and $K$ 
can be parametrized in terms of form factors as follows: 
\begin{equation}
<K(p^\prime)|{\bar s} \gamma_\mu b |B(p)>=(p+p^\prime)_\mu F_1(q^2) 
+{M_B^2-M_K^2 \over q^2} q_\mu \left (F_0(q^2)-F_1(q^2)\right ) 
\label{f0} 
\end{equation}
\noindent ( $q=p-p^\prime$, $F_1(0)=F_0(0)$) and
\begin{equation}
<K(p^\prime)|{\bar s}\; i\;  \sigma_{\mu \nu} q^\nu b |B(p)>=
\Big[(p+p^\prime)_\mu q^2 -(M_B^2-M_K^2)q_\mu\Big] \; {F_T(q^2) \over M_B+M_K}  
\hskip 3 pt . \label{ft} 
\end{equation}
\noindent 
The heavy-to-light meson form factors
$F_1$ and $F_0$ appear in the calculation of two-body nonleptonic
$B \to K X$ decays, if the factorization approximation is adopted; neglecting 
$SU(3)_F$ breaking effects, they govern the semileptonic decay $B 
\to \pi \ell {\bar \nu}$. 

$F_1$ and $F_0$ have already been studied  
by three-point QCD sum rules \cite{ball,pav1}. In the following 
we describe in detail the calculation of $F_T$;
for the sake of completeness, we also report the results for
$F_1(q^2)$ and $F_0(q^2)$ using a unique set of parameters and adopting 
a coherent 
numerical procedure, in order to have at our disposal
a consistent set of form factors.

To compute $F_T$ within the QCD sum rule  approach 
we consider the three-point correlator \cite{ioffe}

\begin{equation}
\Pi_{\alpha \mu \nu} (p,p^\prime,q)= i^2 \int dx dy \;
e^{(i p'\cdot y - i p\cdot x)}  
<0|T[J_\alpha^K(y) J_{\mu \nu}(0) 
J_5^B(x)]|0>  \label{correl} \end{equation}
\noindent 
of the flavour changing quark current
$J_{\mu \nu}={\bar s} i \sigma_{\mu \nu} b$
and of two currents $J_\alpha^K(y)$ and $J_5^B(x)$ with the $K$ and $B$ quantum 
numbers, respectively:
$J_\alpha^K(y)={\bar q}(y) \gamma_\alpha \gamma_5 s(y)$ and 
$J_5^B(x)={\bar b}(x) i \gamma_5 q(x)$. 
The correlator $\Pi_{\alpha \mu \nu}$ can be expanded in  a set of
independent Lorentz structures:
\begin{equation}
\Pi_{\alpha \mu \nu}=
i  p^\prime_\alpha (p_\mu p^\prime_\nu - p_\nu p^\prime_\mu) \Pi  \; + \;
i \sum_n a_{\alpha \mu \nu}^{(n)}  \Pi^{(n)} 
\label{pi} \end{equation}
\noindent 
where $\Pi$ and $\Pi^{(n)}$ are functions of $p^2$, $p'^2$ and $q^2$,
and $a_{\alpha \mu \nu}^{(n)}$ are other tensors set up  using the vectors
$p$ and  $p^\prime$ and the metric tensor $g_{\mu\nu}$.  

Let us consider $\Pi$. 
To incorporate the quark-hadron duality, on which the 
QCD sum rule approach is based, we 
write down  for $\Pi(p^2,p^{\prime 2},q^2)$ a dispersive representation:
\begin{equation}
\Pi(p^2,p'^2,q^2)= {1 \over \pi^2} \int_{m_b^2}^{+ \infty} ds 
\int_{m_s^2}^{+ \infty} ds^\prime \;
{\rho(s,s^\prime,q^2) 
\over (s-p^2)(s^\prime -p^{\prime 2})}
\; + \; subtractions \;  \label{dr} 
\end{equation}
\noindent in the variables $p^2$ and $p^{\prime 2}$ corresponding to the $B$ 
and $K$ channel, respectively. 
 In the region of low values of $s, \; s^\prime$
the physical spectral density $\rho(s, s^\prime, q^2)$ contains a double 
$\delta$-function term corresponding to the transition $B \to K$,
and therefore the function $\Pi$ can be 
written as

\begin{equation}
\Pi =  {R \over (M_B^2-p^2)(M_K^2-p^{\prime 2})} + 
 {1 \over \pi^2} \int_{D} ds ds^\prime {\rho^{had}(s,s^\prime,q^2) 
\over (s-p^2)(s^\prime -p^{\prime 2})} \hskip 3 pt , \label{pi5} 
\end{equation}
\noindent where the residue $R$ 
is given in terms of the form factor $F_T(q^2)$ and of the 
leptonic constants $f_K$ and $f_B$, defined by the matrix elements 
$<0|{\bar q} \gamma_\mu \gamma_5 s|K(p^\prime)>=i f_K p^\prime_\mu$ 
and $<0|{\bar q}i \gamma_5 b |B(p)>=f_B M_B^2/m_b$ (we put $m_q=0$): 
$R= H  F_T(q^2)$ with 
$\displaystyle H= -2 f_K f_B M^2_B/m_b(M_B + M_K)$.
The integration domain $D$ in (\ref{pi5}), where higher resonances 
with the same $B$ and $K$ quantum numbers
contribute to the spectral density $\rho$, starts from two effective thresholds
$s_0$ and $s'_0$.

Also the perturbative contribution to $\Pi$, computed  for
$p^2 \to - \infty$ and  $p^{\prime 2} \to - \infty$, 
can be written as eq.(\ref{dr}). Moreover, considering  the first power 
corrections of the Operator Product Expansion 
of the correlator (\ref{correl}) we 
get the following representation:
\begin{equation}
\Pi(p^2, p'^2, q^2)={1 \over \pi^2} \int_{m_b^2}^{+ \infty} ds 
\int_{m_s^2}^{+ \infty} ds^\prime 
{\rho^{QCD}(s,s^\prime,q^2) 
\over (s-p^2)(s^\prime -p^{\prime 2})}+d_3<{\bar q}q>+d_5 <{\bar q} \sigma G q> 
 +... \hskip 3 pt \; \; . \label{pi5qcd} \end{equation}
\noindent 
$\rho^{QCD}(s,s^\prime, q^2)$ is the perturbative spectral 
function; the two other terms in (\ref{pi5qcd}), expressed as a combination of 
vacuum expectation values of quark and gluon gauge-invariant operators of 
dimension 3 and 5, respectively:
 $<{\bar q}q>$ and $<{\bar q}\sigma G q>=\displaystyle <g_s {\bar q} 
\sigma^{\mu \nu} G_{\mu \nu}^a {\lambda^a \over 2} q>$,  
parametrize the lowest order power corrections. 
The expressions for $\rho^{QCD}$ and $d_5$ can 
be found in Appendix A, eqs.(\ref{rho_ft})-(\ref{d5}); in this particular case
 $d_3$ vanishes.

We now invoke the quark-hadron duality, i.e. we assume that the physical 
and the 
perturbative spectral densities are dual to each other, giving  the same 
result when integrated over an appropriate interval. Assuming duality in the 
region $D$ of the hadronic continuum
\begin{equation}
\int_{D} ds ds^\prime \big\{ \rho^{had}(s,s^\prime,q^2) - 
\rho^{QCD}(s,s^\prime,q^2) \big\} = 0 \label{dual}
\end{equation}
we derive the sum rule for $F_T$:
\begin{equation}
{H F_T(q^2) \over (M^2_B - p^2)(M^2_K - p'^2)}
={1 \over \pi^2} \int_{D'} ds ds^\prime {\rho^{QCD}(s,s^\prime,q^2) 
\over (s-p^2)(s^\prime -p^{\prime 2})}+d_3<{\bar q}q>+d_5 <{\bar q} \sigma G q> 
+... \hskip 3 pt , \label{sumr} \end{equation}
\noindent where $D'$ is 
the region corresponding to the low-lying $B$ and $K$ states:
$m_b^2 \le s \le s_0$, 
$s'_-(s) \le s' \le s'_+(s)$ with 
$\displaystyle s'_\pm(s)= m_s^2 +{(s-m_b^2) \over 2 m_b^2} \left [(m_b^2 +
m_s^2 -q^2) \pm \sqrt{(m_b^2 + m_s^2 -q^2)^2 -4 m_b^2 m_s^2 }\right ]$ and 
$s'_+ \le s'_0$.
The effective thresholds $s_0$ and $s^\prime_0$ can be fixed from the QCD sum 
rule analysis of two-point functions in the $b$ and $s$ channels. We get 
$s_0$ from the calculation of $f_B$, and $s^\prime_0$ from the expected
mass of the first radial excitation of the kaon.

An improvement of the expression in (\ref{sumr}) can be obtained by applying to 
the left and right hand sides 
 the SVZ-Borel transform, defined by
\begin{equation}
{\cal B}_{M^2} {1 \over (m^2 - p^2)^n} = {1 \over (n-1)!} 
{e^{- m^2 /M^2} \over (M^2)^n} \; \; \; , 
\end{equation}
both in the variables $-p^2$ and $-p'^2$;
$M^2$ is a new (Borel) parameter. This operation has the advantage that 
the convergence of the power series is
improved by factorials; moreover, 
for low values of $M^2$ and $M^{\prime 2}$ the possible contribution of 
higher states in eq.(\ref{sumr}) is exponentially suppressed.
The resulting Borel transformed sum rule for $F_T$ reads 
\begin{eqnarray}
H F_T(q^2) e^{-M^2_B/M^2 -M^2_K/M'^2}
&=&{1 \over \pi^2} \int_{D'} ds ds^\prime \rho^{QCD}(s,s^\prime,q^2)
e^{-s/M^2 -s'/M'^2} \nonumber \\
&+& \Big[ {\tilde d}_3<{\bar q}q>+{\tilde d}_5 <{\bar q} \sigma G q> \Big] 
e^{-m^2_b/M^2 -m^2_s/M'^2} \; .
 \label{sr} 
\end{eqnarray}

From eq.(\ref{sr}) the form factor $F_T(q^2)$ can be derived, once the value of 
the Borel parameters $M^2$ and $M'^2$ is fixed. This can be done observing 
that, since $M^2$ and $M'^2$ are unphysical quantities, $F_T$ 
must be independent on them (stability region of the sum rule); moreover, the 
values of $M^2$ and $M'^2$ should allow a hierarchical structure in 
the series of the power correction, and a suppression of the contribution of 
the continuum in the hadronic side of the sum rule.

In our numerical analysis we use 
the values for the quark condensates (at a 
renormalization scale $\mu \simeq 1 \; GeV$) \cite{ioffe}:
\begin{eqnarray}
<\bar q q> & = & (- 230 \; MeV)^3 \nonumber \\
<\bar q g_s \sigma^{\mu \nu} G^a_{\mu \nu} {\lambda^a \over 2 } q> & = &
m_0^2 \; <\bar q q> \label{cond}\end{eqnarray}
\noindent with $m_0^2=0.8 \; GeV^2$. Notice that the numerical results do not 
change sensitively if the condensates 
are evaluated at higher 
scales using the leading-log approximation for their  anomalous dimension. 

As for the quark masses and leptonic constants, we use:
$m_s=0.175 \; GeV$, $ m_b=4.6  \; GeV$, 
$f_K=0.16 \; GeV$ and $f_B =0.18 \; GeV$. 
The thresholds $s_0$ and $s'_0$ are chosen in the range:
$s_0=(33-36)\; GeV^2$ and $s'_0=(1.4-1.6) \; GeV^2$, with the Borel parameters 
kept fixed to the values $M^2=8 \; GeV^2$ and $M'^2=2 \; GeV^2$.
  
Putting these parameters in eq. (\ref{sr}) we obtain the form factor $F_T$  
depicted in fig.1, where  
the different curves 
correspond to different choices of the thresholds $s_0$ and $s_0^\prime$. 
In the sum rule, the perturbative term is a factor of 4$-$5 times larger than 
the $D=5$ contribution, and
 the integral of the spectral function over the region $D'$ gives more 
than $60 \%$ of the result of the integration over the whole region of the 
dispersion relation (\ref{dr}).
The duality window, where the results become independent of the Borel 
parameters $M^2$ and $M^{\prime 2}$, starts at
$M^2 \simeq 7 \; GeV^2$ and $M^{\prime 2}\simeq 1.7 \; GeV^2$; varying
$M^2$ in the range $7 - 9 \; GeV^2$ and
$M^{\prime 2}$ in the range $1.7 - 2.5 \; GeV^2$ the results change within
the bounds provided by the different curves depicted in fig.1.

The same analysis can be applied to 
the form factors $F_1$ and $F_0$ 
using the flavour-changing
vector current $J_\mu={\bar s} \gamma_\mu b$ in the correlator
(\ref{correl}) and studying the projection 
 $q^\mu \Pi_{\alpha \mu}$ to derive $F_0$.
We report in Appendix A the relevant quantities appearing in the sum rules for 
$F_1$ and $F_0$; the difference with respect to \cite{dosch},
 as far as $F_1$ is 
concerned, is that we keep all terms proportional to powers of the strange 
quark mass $m_s$. 
In the calculation of
both the form factors, the contribution of the perturbative term and of
the $D=3$ term have 
comparable size, whereas the $D=5$ term is one order of magnitude smaller; 
the contribution of the resonance in the hadronic side of the rule is nearly 
equal to the contribution of the continuum. We obtain the form factors 
$F_1(q^2)$ and $F_0(q^2)$ depicted in fig.1.
Also in this case the Borel parameters can be varied 
in the range $M^2= 7 - 9 \; GeV^2$ and
$M^{\prime 2}= 1.7 - 2.5 \; GeV^2$; the results change within the region
corresponding to the different curves depicted in fig.1 for each form factor.

We observe a different $q^2$ dependence for the various form factors. 
In the range of $q^2$ we are considering
($0 \le q^2 \le 13-15 \; GeV^2$) $F_1$ follows a
simple pole formula:
\begin{equation}
\displaystyle F(q^2)={F(0) \over 1 - \displaystyle {q^2 \over M_P^2}} 
\hskip 3 pt
\label{polo} 
\end{equation} 
\noindent with $F_1(0)=0.25 \pm 0.03$ and 
$M_{P_1} \simeq 5 \; GeV$. 
A fit to the formula (\ref{polo}) for $F_0$ gives the result 
$M_{P_0}\simeq 7 \; GeV$. The same formula, applied to $F_T$ would give
$F_T \simeq-0.14 $ and $M_P \simeq 4.5 \; GeV$.
Therefore, only the dependence of the form factor
$F_1(q^2)$ does not contradict
the polar behaviour dominated by $B_s^*$, which is the 
nearest singularity in the $t-$ channel, as we would expect
 by invoking the vector meson dominance (VMD) ansatz. The form factor 
$F_0$ increases softly with $q^2$
and, as already observed in  \cite{pav1},
the fitted mass of the pole is larger than the
expected mass of the physical singularity, in this case 
the $J^{P} =0^+$ $b \bar s$ state.
As for $F_T$, the VMD ansatz would predict a polar dependence, with the pole 
represented by $B^*_s$; on the other hand, 
 we observe that $F_T$ can be related to $F_1$ and 
$F_0$  by an  identity obtained by the equation of motion:
\begin{equation}
F_T(q^2)= (M_B + M_K) (m_b + m_s) { F_0(q^2) - F_1(q^2) \over q^2} 
\hskip 3 pt ; \label{wi}
\end{equation}
\noindent eq.(\ref{wi}) is in agreement
with the computed form factor $F_T$ displayed in fig. 1, 
and therefore we can use the double pole model:
\begin{equation}
\displaystyle
F_T(q^2)={F_T(0) \over \left (1 - \displaystyle {q^2 \over 
M_{P_1}^2}\right )
\left (1 - \displaystyle {q^2 \over M_{P_0}^2}\right )}
 \hskip 3 pt \label{dipolo} 
\end{equation}
\noindent 
with $F_T(0)=-0.14 \pm 0.03$ and  $M_{P_1}$ and $M_{P_0}$  given by 
the fitted values of the mass 
of the poles of $F_1$ and $F_0$, respectively. \par
It is interesting to observe that information 
on the possible form  of the $q^2$ 
dependence of the form factors can be derived by studying 
the limit $m_b \to \infty$. In this limit, at the zero recoil point 
where the 
kaon is at rest in the $B$ meson rest frame, it is 
straightforward to show that  the parametric dependence of the form factors
on the heavy meson mass $M_B$ is given by: 
$F_1(q^2_{max}) \sim \sqrt{M_B}$ and 
$F_0(q^2_{max}) \sim 1/\sqrt{M_B}$ \cite{IW}. 
Both these scaling laws are 
compatible with the constraint $F_1(0)=F_0(0)$ and with a multipolar functional 
dependence 
\begin{equation}
\displaystyle
F_i(q^2)={F_i(0) \over 
\left ( 1-\displaystyle {q^2 \over M_{P_i}^2} \right)^{n_i}} \label{fi} 
\end{equation}
\noindent if $n_1=n_0+1$. Thus, in 
the limit $m_b \to \infty$, to a polar  form factor $F_1(q^2)$ 
corresponds a nearly 
constant form factor $F_0(q^2)$. The outcome of QCD sum rules is in 
agreement with this observation \cite{sant}; the observed increasing 
of $F_0$ would be due to subleading terms contributing at finite $m_b$. 

Let us now compare our results with the outcome of different QCD based 
approaches. In the 
channel $B \to \pi$ the form factor $F_1$ has been computed by light-cone 
sum rules \cite{belyaev}, with numerical results in  agreement, at finite 
b-quark mass, with the outcome of three point function sum rules.

As for lattice 
QCD, both $F_1$ and $F_0$ have been computed at large $q^2$
\cite{bur}, and data show that 
$F_0$ has a flat dependence on the momentum transfer, whereas $F_1$
increases with $q^2$. 

The full set of form factors $F_1$, $F_0$ and $F_T$ by these other methods is 
still missing; the complete comparison of our results  with such 
different approaches could help in understanding the drawbacks and the 
advantages of the various methods; this
would shed light on the issue of decays such as 
$B \to \pi \ell \nu$ that  are of
interest as far as the measurement of $V_{ub}$ is concerned.

\section{Form factors of $B \to K^* \ell^+ \ell^-$}
The form factors parametrizing the hadronic matrix elements of the transition 
$B \to K^* \ell^+ \ell^-$ can also be computed by
QCD sum rules by considering a three-point correlator with the interpolating 
current for $K^*$ represented by the vector current
$J_\alpha^{K^*}(y)={\bar q}(y) \gamma_\alpha s(y)$. Let us define
the $B \to K^*$ matrix elements:
\begin{eqnarray}
<K^*(p^\prime,\epsilon)|{\bar s} \gamma_\mu (1-\gamma_5) b |B(p)>&=&
\epsilon_{\mu \nu \alpha \beta} \epsilon^{* \nu} p^\alpha p^{\prime \beta}
{ 2 V(q^2) \over M_B + M_{K^*}}  \nonumber \\
&-& i \left [ \epsilon^*_\mu (M_B + M_{K^*}) A_1(q^2) -
(\epsilon^* \cdot q) (p+p')_\mu  {A_2(q^2) \over (M_B + M_{K^*}) } 
\right. \nonumber \\ 
&-& \left. (\epsilon^* \cdot q) {2 M_{K^*} \over q^2}   
\big(A_3(q^2) - A_0(q^2)\big) 
q_\mu \right ]
\label{a1}
\end{eqnarray}
\noindent and
\begin{eqnarray}
<K^*(p^\prime,\epsilon)|{\bar s} \sigma_{\mu \nu} q^\nu 
{(1+\gamma_5) \over 2} b |B(p)>&=&
i \epsilon_{\mu \nu \alpha \beta} \epsilon^{* \nu} p^\alpha p^{\prime \beta}
\; 2 \; T_1(q^2)  + \nonumber \\
&+&  \Big[ \epsilon^*_\mu (M_B^2 - M^2_{K^*})  -
(\epsilon^* \cdot q) (p+p')_\mu \Big] \; T_2(q^2) \nonumber \\ 
&+& (\epsilon^* \cdot q) 
\left [ q_\mu - {q^2 \over M_B^2 - M^2_{K^*}} (p + p')_\mu \right ] 
\; T_3(q^2)  \; .
\label{t1}
\end{eqnarray}
\noindent
$A_3$ can be written as a linear combination of $A_1$  and $A_2$:
\begin{equation}
A_3(q^2) = {M_B + M_{K^*} \over 2 M_{K^*}}  A_1(q^2) - 
{M_B - M_{K^*} \over 2 M_{K^*}}  A_2(q^2)  
\end{equation}
with the condition
$ A_3(0) = A_0(0)$. The identity
$\displaystyle \sigma_{\mu \nu} \gamma_5 = - {i \over 2} 
 \epsilon_{\mu \nu \alpha \beta} \sigma^{\alpha \beta}$
($ \epsilon_{0 1 2 3}=+1$) implies that $T_1(0) = T_2(0)$.

The form factors $T_1(q^2)$ and $T_2(q^2)$  can be derived by the correlator

\begin{equation}
\tilde \Pi_{\alpha \mu} (p,p^\prime,q)= i^2 \int dx dy \;
e^{(i p'\cdot y - i p\cdot x)} \; 
<0|T[J_\alpha^{K^*}(y) \tilde J_\mu(0) 
J_5^B(x)]|0> \hskip 3 pt , \label{correl1} \end{equation}
\noindent 
with 
$\displaystyle\tilde J_\mu={\bar s}  \sigma_{\mu \nu} {1 + \gamma_5
\over 2} q^\nu b$. 
Expanding $\tilde \Pi_{\alpha \mu}$ in Lorentz independent structures
\begin{equation}
\tilde \Pi_{\alpha \mu}=
i \epsilon_{  \alpha \mu \rho \beta} p^\rho p^{\prime \beta} \tilde \Pi_1
+ g_{\alpha \mu} \tilde \Pi_2 + other \; structures \; in \; p , \; p^\prime 
\label{pit} \end{equation}
we get $T_1$ and $T_2$ from $\tilde \Pi_1$ and $\tilde \Pi_2$, respectively.
The sum rules have the same structure  of eqs. (\ref{sumr}), (\ref{sr}), with 
the perturbative spectral functions 
$\rho(s,s^\prime,q^2)$ and the power corrections 
$d_3$ and $d_5$ reported in Appendix B. The only difference with 
respect to the kaon case is the value of the $K^*$ leptonic 
constant, defined by the matrix element 
$<0|{\bar q}\gamma_\mu s|K^*(p,\epsilon)>=f_{K^*} M_{K^*} \epsilon_\mu$, 
with $f_{K^*}=216 \; MeV$. 

In fig.2 we depict the form factors  $T_1(q^2)$ and $T_2(q^2)$ obtained  
choosing the threshold  
$s_0^\prime$ in the range $1.6-1.8 \; GeV^2$ and the other parameters as in 
the previous section.
In the sum rule for both the form factors the perturbative term does not 
dominate over the non-perturbative ones: at $q^2=0$ it represents
$30\%$ of the quark condensate contribution, and is nearly equal to the $D=5$ 
term. However, it rapidly increases with the momentum transfer, and at
$q^2=15 \; GeV^2$ it is equal to the contribution of the $D=3$ term, whereas
the $D=5$ contribution is an order of magnitude smaller.

Concerning the form factor $T_3$, we observe that
it contributes, together with $T_1$ and $T_2$, 
to other invariant functions in (\ref{pit}) and, in principle, it also could  
be obtained by a sum rule. However, since it can be related
to $A_1$, $A_2$ and $A_0$ by applying the equation of motion:
\begin{equation}
T_3(q^2)=M_{K^*} (m_b-m_s) {A_3(q^2)-A_0(q^2) \over q^2} \hskip 3 pt  
\label{t3} 
\end{equation}
we prefer to use this expression to determine it, considering 
that this procedure is successful for $F_T(q^2)$.

The form factors $V$ and $A_i$ can be obtained by studying the correlator
(\ref{correl1}) with a vector 
$J^V_\mu= {\bar s} \gamma_\mu b$ and an axial
$J^A_\mu= {\bar s} \gamma_\mu \gamma_5 b$ flavour changing current, considering 
the projection $q^\mu J^A_\mu$ to derive $A_0$. We collect in Appendix B
the complete expressions appearing in the relevant sum rules 
for all the form factors, excluding $A_0$, 
whose expressions can be found in \cite{noi1}; also in this case the difference 
with respect to \cite{dosch} is that we include all powers of the strange quark 
mass.
 
Using our set of parameters we get
$V(q^2)$, $A_1(q^2)$, $A_2(q^2)$ and $A_0(q^2)$ 
depicted in fig.3, and, using (\ref{t3}), 
the form factor $T_3$ in fig.2.

As it happens for $T_1$ and $T_2$,
also in the sum rules for $V$, $A_1$ and $A_2$ the perturbative term, at 
$q^2=0$, is smaller than the $D=3$ contribution;  the relative weights 
of the various contributions change with the momentum transfer, and at $q^2=15 
\; GeV^2$ the $D=0$ and $D=3$ terms have comparable size.
As it happens for the $B \to K$ form factors, 
the chosen values of $M^2$ and $M^{\prime 2}$,
$M^2=8 \; GeV^2$ and $M^{\prime 2}=2 \; GeV^2$, are within the duality window 
where the results are independent of the Borel parameters.
Also in this case, varying $M^2$ and $M^{\prime 2}$ in the ranges
$M^2=7 - 9 \; GeV^2$ and $M^{\prime 2}=1.7 - 2.5  \; GeV^2$, the final results
change within the same uncertainty coming from the 
variation of the continuum threshold.

Considering the results displayed in figs.2 and 3, we collect  
the form factors $T_i$, $V$ and $A_i$
in three sets, according to their functional dependence on 
the momentum transfer. In the first set we include
$T_1$, $V$ and $A_0$, that display a sharp increasing with $q^2$. 
It is possible 
to fit them with a polar $q^2$ dependence eq.(\ref{polo}) 
(as observed also in \cite{ball,noi1}) with:
$T_1(0)=0.19 \pm 0.03$ and 
$M_P\simeq 5.3 \; GeV$,
$V(0)=0.47 \pm 0.03$ and $M_P\simeq 5 \; GeV$,
$A_0(0)=0.30 \pm 0.03$ and $M_P\simeq 4.8 \; GeV$
(the difference with respect to the value 
$T_1(0)=0.17 \pm 0.03$ in ref.\cite{noi}
is due to the effect of the strange quark mass, that here has been included).

The error on the mass of the pole is correlated to the 
error on the form factor at $q^2=0$, and it can be estimated of the order of 
$200-300 \; MeV$. The relevant result is that the masses of the poles 
are not far from the values expected by the dominance of the nearest 
singularity in the $t-$ channel:  $M_P= M_{B^*_s}$ for $T_1$ and $V$,
$M_P= M_{B_s}$ for $A_0$.
We stress that the fit is performed in a range of values of
$q^2$ where the QCD calculation can be meaningfully carried out, therefore
large momentum transferred $[q^2 > 15 \; GeV^2]$ are not taken into 
account.

In the second set of form factors we include
$T_2$, $T_3$ and $A_1$. They softly decrease with $q^2$: 
$F_i(q^2)=F_i(0) (1 + \beta q^2)$, with
$T_2(0)=T_1(0)$ and $\beta= - 0.02 \; GeV^{-2}$,
$T_3(0)=-0.7$ and $\beta= 0.005 \; GeV^{-2}$,
$A_1(0)=0.37 \pm 0.03$ and $\beta= - 0.023 \; GeV^{-2}$ with the error on 
$\beta$ at the level of $10\%$.
The dependence of $T_3$ is related to $A_1$, $A_2$ and $A_0$.

The last form factor, $A_2$, linearly increases with $q^2$:
$A_2(0)=0.40 \pm 0.03$ and $\beta= 0.034 \; GeV^{-2}$. A fit to a polar 
dependence for this form factor would  give $M_P \ge 7 \; GeV$
for the mass of the pole. 

The parameters of all the form factors are collected in Table II.
Albeit the form factors have been computed in a well defined range of momentum 
transfer, once their functional $q^2$ dependence has been fitted and the 
parameters determined, we extrapolate them up to $q^2_{max}$. This procedure 
cannot be avoided within the method of QCD sum rules, where large positive 
values of $q^2$ are not accessible since there is a region where the 
distance between the points $x$, $y$ and $0$ in the  
correlator, which is the initial ingredient of this approach,
 is large, and  therefore the standard OPE cannot be used; this is
shown by the occurrence of singularities in the correlator when $q^2$ 
is close to $q^2_{max}$. 

As for the computed dependence on the momentum transfer, is worth reminding 
that deviations from the VMD expectations 
for the form factors $A_1$ and $A_2$ 
have been already observed in the literature, first in the 
 $D \to K^* \ell \nu$ \cite{dosch} channel and then for $B \to \rho \ell \nu$
\cite{ball}. 
Here we find a kind of common feature, i.e. all form factors deviating from the 
polar dependence (excluding $F_T$) seem to depend linearly 
on the momentum transfer, with small (positive or negative) slopes. 

It is interesting that also for 
$T_1(q^2)$ and $T_2(q^2)$ we can use
the argument developed in the previous section 
concerning the limit $m_b \to \infty$: since $T_1(q^2_{max})
\sim \sqrt{M_B}$ and $\displaystyle T_2(q^2_{max})\sim 1/\sqrt{M_B}$, 
the constraint $T_1(0)=T_2(0)$ can be fulfilled 
by a multipolar $q^2$ dependence if $n_1=n_2+1$ in eq. (\ref{fi}). 

At zero momentum transfer our results numerically 
agree with those obtained by the method of light-cone sum rules
\cite{simma}, within the errors and taking into account the different 
choices of the input parameters. In \cite{simma} it has also been observed that
 $T_1$, $V$ and $A_1$ have  different functional dependencies  on $q^2$;
the difference with respect to our case is that 
the slopes are larger than those obtained from three-point sum rules; in 
particular, the form factor $A_1$ increases with $q^2$. The origin of this
discrepancy should be investigated.

The form factors $T_1$ and $T_2$ have been computed by lattice 
QCD \cite{lat,lat1} near the point of zero recoil and for the 
mass of the heavy quark 
smaller than $m_b$, due to the finite size of the available lattices; therefore,
the results at $q^2=0$ and for a realistic value of $m_b$ are obtained after 
an extrapolation in the momentum transfer and in the heavy quark mass.
Also in this case, in the region of 
large values of $q^2$, 
the form factor $T_1$ increases rapidly with the momentum 
transfer, whereas $T_2$ is quite flat.  
As for the analytic $q^2$ behaviour obtained from lattice 
calculations, it seems to us that larger lattices are needed to enlarge
the range of momentum transfer where the measurements can be performed,
in order to clearly disentangle different possible dependencies of $T_1$ and 
$T_2$ (e.g., dipole {\it versus} pole or pole {\it versus} constant).

\section{Relations between rare and semileptonic $B$ decay form factors}
In the limit $m_b \to \infty$ 
Isgur and Wise  \cite{IW} and Burdman and Donoghue 
\cite{BD} have derived exact relations between the form factors $F_T$, 
$T_i$ in eqs.(\ref{ft}), (\ref{t1}) and the form factors $F_i$, $V$, $A_i$
in eqs.(\ref{f0}), (\ref{a1}). 
These relations can be easily worked out
observing that,  in the effective theory where the $b$-quark mass is taken 
to the infinity, the equation $\gamma^0 b =b$ is fulfilled
in the rest frame of the ${B}$ meson.  

In our parametrization 
such relations can be written as follows, near the point of zero recoil ($q^2 
\simeq q^2_{max}=(M_B-M_{K^{(*)}})^2$):
\begin{equation}
F_T(q^2)=-{M_B+M_K \over 2 M_B} \left [ F_1(q^2)-(M_B^2-M_K^2)
{F_0(q^2)-F_1(q^2) \over q^2} \right ] 
\label{is_ft} 
\end{equation}
\begin{equation}
T_1(q^2)={M_B+M_{K^*} \over 4 M_B} A_1(q^2)+{M_B^2-M_K^{*2}+q^2 \over 4 M_B
(M_B+M_{K^*})} 
V(q^2)  \label{is_t1} \end{equation}
\begin{equation}
T_2(q^2)={(M_B^2-M_{K^*}^2+q^2) \over 4 M_B (M_B-M_{K^*})} A_1(q^2) +
{\lambda(M_B^2,M_{K^*}^2,q^2) \over 
4 M_B (M_B-M_{K^*})(M_B+M_{K^*})^2} V(q^2)
\label{is_t2} \end{equation}
\begin{equation}
T_3(q^2)=-{M_B^2+3M_{K^*}^2-q^2 \over 4 M_B  (M_B+M_{K^*}) } V(q^2) +
{M_{K^*} \over 2 M_B} A_3(q^2) + 
{M_{K^*} (M_B^2-M_{K^*}^2) \over 2 M_B} {A_3(q^2) - A_0(q^2) \over q^2} 
\label{is_t3} 
\end{equation}
\noindent where $\lambda$ is the triangular function.

It has been argued by several authors  that the relations 
(\ref{is_ft})-(\ref{is_t3})
could also be valid 
at low values of $q^2$ \cite{BD}, although a general proof has not been found
in support of this hypothesis. 

Using the form factors computed by QCD sum rules 
in the previous Sections,  it is possible to
check  eqs.(\ref{is_ft})-(\ref{is_t3}). In fig.4 we plot the ratio
${\cal R}=F/F^{IW}$ in the case of 
$F_T$, $T_1$, and $T_2$, as a function of $q^2$, in the range of momentum 
transfer where the calculation has been carried out. 
 We observe that the relations 
between the various form factors are verified at different level of accuracy. 

In the case of $F_T$ the ratio $\cal R$ differs from unity at the level of 
$25-30 \%$, including the uncertainty coming from 
 the errors of the various parameters. 
In particular, at $q^2=0$ we have $F_T/F^{IW}_T=0.7 \pm 0.1$.
The situation is different for the ratios concerning 
$T_1$ and $T_2$, that differ from unity at the level of  
$10-20 \%$:
at $q^2=0$ we have $T_1/T^{IW}_1=0.94 \pm 0.05$ and
$T_2/T^{IW}_2=1.12 \pm 0.05$.

These results support the argument put forward in \cite{noi} on the
validity of the Isgur-Wise relations, in the limit 
$m_b \to \infty$ also at small values of $q^2$; they also can be well compared 
to the outcome of light-cone sum rules, obtained for $T_1$ at a finite
$m_b$ \cite{simma}. The conclusion is that the 
$b$ quark is near to the mass shell also when the recoil of the light hadron is 
large with respect to $m_b$, with $1/m_b$ corrections that do not appear to 
overwhelm the effect. 

The relations 
(\ref{is_ft})-(\ref{is_t3}) could be used to perform a model independent 
analysis of the decays $B \to K^{(*)} \ell^+ \ell^-$ 
employing experimental information (when available) on the form factors of
the semileptonic transition $B \to \rho \ell \nu$  
\cite{burdman}.
In particular, since (\ref{is_ft})-(\ref{is_t3}) are valid on general grounds
in the large $q^2$ region, it has been proposed to perform the  analysis 
in the 
range of large invariant mass of the lepton pair, e.g. 
$M_{\ell^+ \ell^-} \ge 4 \; GeV$.
 
Albeit in principle correct, we feel
that, from the experimental point of view, 
the procedure of extracting the semileptonic $B \to \rho$ form factors 
near zero recoil will be rather difficult,
with large uncertainties in the final result.
The problem is not avoided by the possible choice of using the form
factors of the semileptonic transition $D \to K^* \ell^+ \nu$, and then
rescaling them according to the their leading dependence on the heavy mass, i.e.
$\displaystyle {V^{B \rho}(q^2_{max}) \over V^{D
K^*}(q^2_{max})}=\sqrt{{M_B \over M_D}}$,  etc. (neglecting $SU(3)_F $  and
$\alpha_s$ corrections). As a matter of fact, in such procedure the
next-to-leading mass corrections could be large and not under control.
Finally, as we shall see in the next section, the differential branching
ratios of $B \to K^{(*)} \ell^+ \ell^-$ at large $q^2$ are small,
and therefore the experimental errors are expected to be sizeable.
For this reason we prefer to propose an analysis of the decay
extended to the full range of $q^2$, using hadronic quantities 
determined in a well defined theoretical framework. The dependence
on the computational scheme will be reduced  once the different form factors 
have been computed by different QCD calculations, and the whole information 
collected in a unique set of form factors.

\section{Decay $B \to K \ell^+ \ell^-$}

We can now compute the invariant mass squared distribution of the lepton pair 
in the decay $B \to K \ell^+ \ell^-$:
\begin{eqnarray}
{d \Gamma \over d q^2} && (B \to K \ell^+ \ell^-)={M_B^3 G_F^2 \alpha^2 \over 
1536 \pi^5} |V^*_{ts} V_{tb}|^2 \times \nonumber \\
&&\left\{ \left|C_7\,2 m_b \left (- {F_T(q^2) \over M_B+M_K} \right )+ 
C_9^{eff} F_1(q^2)\right |^2+
\left | C_{10} F_1(q^2)\right |^2 \right\} \times \nonumber \\
&& \left[ \left ( 1-{M_K^2 \over M_B^2} \right )^2 + 
\left ({q^2 \over M_B^2}\right )^2-2
\left(
{q^2 \over M_B^2} \right) \left (1+{M_K^2 \over M_B^2} \right) \right ]^{3/2} 
\label{spettro_k} 
\end{eqnarray}
\noindent 
($q^2=M^2_{\ell^+ \ell^-}$).
The contribution of the operators
$O_7$, $O_9$ and $O_{10}$ is taken into account in the terms proportional to
$C_7$, $C_9$ and $C_{10}$.
The operators $O_1$ and $O_2$ provide a short distance contribution, with a 
loop of charm quarks described by the function $h(x,s)$ $\left
(\displaystyle  x=m_c/m_b,\, s=q^2/m_b^2\right )$
\cite{gri,misiak}: 
\begin{equation}
h(x,s)=-\Bigg[ {4 \over 9} ln x^2 -{8 \over 27} -{16 \over 9} {x^2 \over s} + 
{4\over 9} \sqrt{{4 x^2 \over s}-1} \Bigg(2 + {4 x^2 \over s}\Bigg) 
\displaystyle arctg\left (\displaystyle {4x^2 \over s}-1 \right)^{-1/2} \Bigg] 
\label{h(x,s)} \end{equation}
\noindent if $s < 4 x^2$, and
\begin{equation}
h(x,s)=-\Bigg\{ {4 \over 9} ln x^2 -{8 \over 27} -{16 \over 9} {x^2 \over s} + 
{2\over 9} \sqrt{1-{4 x^2 \over s}} \Bigg(2 + {4 x^2 \over s}\Bigg) 
\Bigg [ ln \left|
{1+  \sqrt{1- 4x^2/ s} \over
1-\sqrt{1-4x^2/s}} \right|  - i \pi \Bigg]
\Bigg\} \label{h(x,s)_1} \end{equation}
\noindent if $s>4 x^2$;
the imaginary part in (\ref{h(x,s)_1}) comes from on-shell charm quarks. 
$O_1$ and $O_2$ also provide a long distance contribution,
related to $c{\bar c}$ 
bound states $(J/\psi, \psi^\prime)$ converting into the lepton pair $\ell^+ 
\ell^-$ \cite{ld,ld1}. 
This contribution can be described in terms of  the $J/\psi$ and 
$\psi^\prime$ leptonic decay constants $<0|{\bar c}\gamma^\mu c|\psi_i
(\epsilon,q)>=\epsilon^\mu f_{\psi_i} M_{\psi_i}$ and of 
the full $J/\psi$ and $\psi^\prime$ decay widths $\Gamma_{\psi_i}$.
We derive $f_{\psi_i}$ 
from the experimental branching ratio $\psi_i \to \ell^+ \ell^-$;  in this way
the whole
contribution of $O_1$ and $O_2$ can be taken into account by modifying the 
coefficient $C_9$ into $C_9^{eff}$:
\begin{equation}
C_9^{eff}=C_9+(3 C_1+C_2)\left [h(x,s) + { k} \; \sum_{i=1}^2
 {\pi \Gamma(\psi_i  \to
 \ell^+ \ell^-) M_{\psi_i} \over q^2-M_{\psi_i}^2+i M_{\psi_i} 
\Gamma_{\psi_i}} \right ] \;. 
\label{c9eff} \end{equation}
\noindent 
If the nonleptonic $B \to K \psi_i$ transition is computed by factorization, the
parameter $k$ is given by $\displaystyle {k}= {3 \over \alpha^2}
{|V^*_{cb} V_{cs}|\over |V^*_{ts} V_{tb}|}$; the sign between the short
distance and the long distance term in (\ref{c9eff}) can be fixed
according to the analyses in ref.\cite{ld1}. In ref.\cite{ali} the value
of $ k$ is appropriately chosen in order to reproduce the quantity: 
\begin{equation}
\left. \begin{array}{c}\end{array}
B(B \to K \ell^+ \ell^-)\right |_{res}
=\sum_{i=1}^2 B(B \to \psi_i K) B(\psi_i \to \ell^+ 
\ell^-) \simeq 7 \; 10^{-5} \;\;\; \cite{hon}. \label{brpsi} \end{equation}
\noindent 
This can be done by choosing $\displaystyle { k} \simeq ( 1.5\,\div\,2 )
\times {3 \over \alpha^2}$. Notice that, since the $J/\psi$ and
$\psi^\prime$ resonances are narrow, their contribution modifies the
dilepton spectrum only in the region close to $M^2_{\ell^+ \ell^-}=
M^2_{J/\psi}, \; M^2_{\psi^\prime}$. 

As input parameters we choose the ratio $\displaystyle m_c/m_b 
=0.27-0.29$ and the value of the CKM matrix element $|V_{ts}| \simeq
0.04$; a different value for $|V_{ts}|$ only modifies the prediction of
the branching ratio, leaving unchanged the shape of the spectrum
\cite{grif}. 

We depict in fig.5 the obtained invariant mass squared
distribution of the lepton pair in $B \to K \ell^+ \ell^-$. In the same figure 
we also plot the spectrum obtained considering only
the short distance contribution,
that gives the branching ratio
(using $\tau_B=1.5 \; 10^{-12} \; sec$ for the $B-$ meson lifetime)
 $B(B \to K \ell^+ \ell^-)|_{sd} 
\simeq 3 \times 10^{-7}\displaystyle \left |V_{ts}/0.04 \right |^2$,
to be compared to the experimental upper limit (obtained excluding
the region near $J/\psi$ and $\psi^\prime$)
$B(B^- \to K^- \mu^+ \mu^-) < 0.9 \times 10^{-5} \;$ 
(at $ 90 \% \; CL$) \cite{cleo,argus}.
The uncertainty coming from the two possible values of $C_9$ in Table I is less 
then $1 \%$ and does not have relevant consequences on the predicted branching
 ratio and on the invariant mass distribution.

From the experimental point of view,
the measurement of the spectrum in fig.5 is a non trivial task;
hopefully, it will be possible to obtain experimental results from 
the future dedicated $e^+ e^-$ colliders. The important point to be stressed 
is that, 
in the distribution depicted in fig.5 the
theoretical uncertainty connected to the hadronic matrix element 
is reduced to a well defined QCD computational scheme
(QCD sum rules), so that in the studies of the effects of
interactions beyond the Standard Model the hadronic uncertainty 
plays no more  a major role.

\section{Decay $B \to K^* \ell^+ \ell^-$}
A great deal of information can be obtained from the channel $B \to K^* \ell^+ 
\ell^-$ investigating, together with the lepton invariant mass distribution, 
also the forward-backward (FB) asymmetry in the dilepton angular
distribution; this may reveal effects beyond the Standard Model that
could not be observed in the analysis of the decay rate. \par 
A FB asymmetry in the dilepton 
angular distribution is a hint on parity violation. Since the decay 
$B \to K^* \ell^+ \ell^-$ proceeds through $\gamma$, $Z$ and $W$ 
intermediate bosons, we expect a different behaviour in the various $q^2$ 
kinematical regions. In the region of low $q^2$, the photon exchange dominates, 
leading to a substantially vector-like parity-conserving interaction; 
 as a consequence, we expect a small asymmetry. On the other hand, when 
$q^2$ is large, the contribution of $Z$ and $W$ exchange diagrams becomes 
important, and the interaction acquires the V-A parity violating structure, 
leading to a large asymmetry. As already observed in ref. \cite{liu} this 
pattern strongly depends on the value of the top quark mass, 
and the penguin diagrams with $Z$ exchange and the $W$ box diagram 
are expected to overwhelm the photon penguin diagram in correspondence to the 
measured $m_t$. Moreover,
since the FB asymmetry is sensitive 
not only to the magnitude of the Wilson coefficients,
 but also to their sign \cite{ali}, it can be used 
to probe the values predicted by the Standard Model.

Let us define $\theta_\ell$ as the angle between the $\ell^+$  direction and 
the $B$ direction in the rest frame of the lepton pair. Since, in the case of 
massless leptons, as we assume,  the amplitude 
can be written as sum of non interfering helicity amplitudes, the double 
differential decay rate reads as follows:
\begin{eqnarray}
{d^2 \Gamma \over dq^2 dcos\theta_\ell}&=&{G_F^2|V_{tb}V_{ts}^*|^2
\alpha^2 \over 2^{13} \pi^5} {\lambda^{1/2}(M_B^2,M_{K^*}^2,q^2) \over M_B^3} 
\times 
\left\{ sin^2 \theta_\ell A_{L} + \right.\nonumber \\
&+& \left. q^2\left [(1+cos \theta_\ell)^2(A^L_+ + A^R_-) +
(1-cos \theta_\ell)^2(A^L_- + A^R_+)\right]\right\} \label{dg2} \end{eqnarray}
\noindent where $A_{L}$ corresponds to a longitudinally polarized $K^*$, 
while $A^{L(R)}_{+(-)}$ represent the contribution from left (right) leptons 
and from $K^*$ with transverse polarization: 
$\displaystyle\epsilon_\pm=\Big(0,{1 \over \sqrt{2}},\pm {i \over
\sqrt{2}},0 \Big)$.\par 
We obtain
\begin{equation}
A_{L}={1 \over M_{K^*}^2} \left\{\left|B_1(M_B^2-M_{K^*}^2-q^2) + B_2
\lambda\right|^2+ \left|D_1(M_B^2-M_{K^*}^2-q^2) + D_2 
\lambda\right|^2\right\} 
\label{along} \end{equation}
\noindent and
\begin{equation}
A^L_\pm=|\lambda^{1/2}(A-C) \mp (B_1-D_1)|^2 \label{left} \end{equation}
\begin{equation}
A^R_\pm=|\lambda^{1/2}(A+C) \mp (B_1+D_1)|^2 \; ,\label{right} \end{equation}
\noindent where $\lambda=\lambda(M_B^2,M_{K^*}^2,q^2)$. The terms
$A$,$C$,$B_1$,$D_1$ contain the short distance coefficients, as 
well as the form factors:
\begin{equation}
A={C_7 \over q^2}\,4\, m_b\, T_1(q^2)+C_9\,{V(q^2) \over M_B + M_{K^*}}
 \label{a}
\end{equation}
\begin{equation}
C= C_{10}\, {V(q^2) \over M_B+M_{K^*}} \label{c} \end{equation}
\begin{equation}
B_1={C_7 \over q^2}\, 4\, m_b\, T_2(q^2)(M_B^2-M_{K^*}^2)+
C_9\, A_1(q^2) ( M_B + M_{K^*}) \label{b1}
\end{equation}
\begin{equation}
B_2=- \left [{C_7 \over q^2}\, 4\, m_b\, \left ( T_2(q^2)+q^2 
{T_3(q^2) \over (M_B^2-M_{K^*}^2)} \right )+
C_9 {A_2(q^2)\over   M_B + M_{K^*}} \right ] \label{b2}
\end{equation}
\begin{equation}
D_1= C_{10}\,A_1(q^2)\,(M_B+M_{K^*}) \label{d1} \end{equation}
\begin{equation}
D_2=- C_{10} \,{A_2(q^2) \over M_B+M_{K^*}} \; . \label{d2} \end{equation}
The FB asymmetry is defined as 
\begin{equation}
A^{FB} (q^2)=\displaystyle
{\displaystyle \int_0^1{d^2 \Gamma \over dq^2 d cos\theta_\ell}dcos\theta_\ell
-\int^0_{-1}{d^2 \Gamma \over dq^2 d cos\theta_\ell}dcos\theta_\ell     
\over\displaystyle
\int_0^1{d^2 \Gamma \over dq^2 d cos\theta_\ell}dcos\theta_\ell
+\int^0_{-1}{d^2 \Gamma \over dq^2 d cos\theta_\ell}dcos\theta_\ell} \; ,
 \label{asim_def} 
\end{equation}
\noindent 
thus we have
\begin{equation}
A^{FB}(q^2)={3\over 4}{ 2 q^2 (A^L_+ +A^R_- -A^L_- -A^R_+) \over A_{L} +
2 q^2 (A^L_- + A^R_+ + A^L_+ + A^R_-)} \; .\label{asim}
\end{equation}
\noindent $A^{FB}(q^2) $ is depicted in fig.6;
it is consistent with the 
prediction of low asymmetry in the small $q^2$ region and high asymmetry 
for large $q^2$. The analysis of the individual shapes of the helicity 
amplitudes (neglecting the long distance contribution)
shows that $A^L_+$ and $A^R_+$ have comparable size, and therefore 
there is a cancellation of their contribution in eq.(\ref{asim}); moreover,
they are small with respect to $A_-^{L,R}$.
In the region of large $M^2_{\ell^+ \ell^-}$,  
$A_-^L$ dominates over $A^R_-$, whereas the situation is reversed
 for low dilepton invariant mass squared, 
and this is the reason of the small positive asymmetry
appearing in fig.6 for $M^2_{\ell^+ \ell^-} \le 3 \; GeV^2$. It is interesting 
to observe that such positive asymmetry depends on $C_7$, and 
that it disappears if  $C_7$ has a reversed sign.

The invariant mass squared 
distribution of the lepton pair is depicted in fig.7, where
the short distance contribution is separately displayed.
The predicted branching ratio is
$B(B \to K^* \ell^+ \ell^-)|_{sd}= 1 \times 10^{-6} |V_{ts}/0.04|^2$,
to be compared to the experimental upper limit:
$B( {\bar B^0} \to K^{*0} \mu^+ \mu^-) < 3.1 \times 10^{-5}$ (CLEO II) and
$B( {\bar B^0} \to K^{*0} \mu^+ \mu^-) < 2.3 \times 10^{-5}$ (UA1) 
(at $ 90 \% \; CL$) obtained excluding the region of the 
resonances $J/\psi$ and $\psi'$ \cite{cleo,ua1}, \cite{hon}. 
Also in this case the uncertainty on $C_9$ does not have relevant consequences. 

The interesting observation is that, for low values of the invariant mass 
squared, the distribution is still sizeable, an effect that could be 
revealed at future $B$-factories such as the Pep-II asymmetric $e^+ e^-$  
collider at SLAC. 

\section{Conclusions}
In this paper we have analyzed some features of the rare decays 
$B \to K \ell^+ \ell^-$ 
and $B \to K^* \ell^+ \ell^-$  within the theoretical framework provided 
by the Standard Model, using an approach based on three point function
QCD sum rules to compute the relevant hadronic matrix elements. 

Albeit QCD sum rules have their own limitations 
(finite number of terms in the Operator Product Expansion of the 
correlators, values of the condensates, validity of the local duality 
assumption), we believe that the obtained results are meaningful 
from the quantitative point of view. 

There is a quite good
agreement with independent QCD 
methods (lattice QCD, light-cone sum rules) for few quantities computed by the 
various approaches. The calculations of the remaining quantities 
($F_0$, $T_i$, $A_0$) by the other two methods is required in order to 
complete the overview on the various results. 

We have used our results to test some relations among the computed form factors 
which hold in the infinite heavy quark limit, but that are expected to hold 
also for low values of $q^2$ and for finite $b$ mass. We have found that the 
different form factors satisfy with different accuracies these relations, which 
can be explained by a different role of the $1/m_b$ corrections.

As for the decays we have 
analyzed in the present paper, within the Standard Model they are expected with 
branching ratios of the order $10^{-7}$ ($B \to K \ell^+ \ell^-$) and 
 $10^{-6}$ ($B \to K^* \ell^+ \ell^-$), with peculiar  shapes of the invariant 
mass of the lepton pair and of the FB asymmetry. Any deviation from the above 
expectations would be interpreted as a signal of deviation from the Standard 
Model. 
Interesting experimental data are therefore expected from current and future 
$e^+ e^-$ colliders 
in this exciting sector of the heavy flavour physics.

\acknowledgments

We thank V.M.Braun, F.Buccella, G.Nardulli and N.Paver for useful discussions.

\newpage
\appendix
\section{{\bf $B \to K \ell^+ \ell^-$} }
The three-point function QCD sum rule for the form factor $F_T(q^2)$ 
in eq.(\ref{ft}) can be derived 
by studying the function $\Pi(p^2, p^{\prime 2}, q^2)$ 
in eq. (\ref{pi}) and using in eqs. (\ref{sumr}), 
(\ref{sr}) the following expressions:
\begin{equation}
H=- 2 f_K f_B {M_B^2 \over m_b}  { 1 \over M_B + M_K} \label{h} 
\end{equation}
\begin{equation}
\rho(s,s^\prime,q^2) = {3 \over 2 \lambda^{3\over 2}} 
\left\{2 \Delta' s - \Delta u + \; {1 \over \lambda} \; 
\left [6 \Delta^{\prime 2} s^2 + 2 \Delta s s' - 6 \Delta \Delta' s u + \Delta^2 u^2
\right]
\right\}
\label{rho_ft}
\end{equation}
\noindent with $\Delta=s - m_b^2$, $\Delta^\prime= s^\prime-m_s^2$, 
$u=s+s^\prime -q^2$, $\lambda=u^2-4 s s^\prime$. 
The coefficients of $D=3$ and $D=5$ vacuum matrix elements are given by:
\begin{equation}
d_3=0
\end{equation}
\begin{equation}
d_5={m_b \over 3 r^2 r^{\prime 2}} 
\label{d5} \end{equation}
\noindent with $r=p^2-m_b^2$, $r^\prime=p^{\prime 2}- m_s^2$. \par
Also $F_1(q^2)$ and $F_0(q^2)$ can be derived by equations analogous 
to (\ref{sumr}), (\ref{sr}).
 The relevant quantities for $F_1(q^2)$ are given by:
\begin{equation}
H= f_K f_B {M_B^2 \over m_b} \label{hf1} \end{equation}
\begin{eqnarray}
\rho(s,s^\prime,q^2)&=&{3 \over 8 \lambda^{3/2}} \Bigg\{m_b \left[2 \Delta 
(u-s^\prime) + \Delta^\prime (u-4s)\right ]+m_s(\Delta u -2 \Delta^\prime s) 
\nonumber \\
&+& {2 m_b \over \lambda}\left[ \Delta^2(3 s^\prime u-2s s^\prime-u^2) + 
\Delta^{\prime 2} (3 s u - 6 s^2) 
+ 2 \Delta \Delta^\prime (3 s u -2 s s^\prime 
-u^2)\right] \Bigg\} \label{rho_f1}  \end{eqnarray}
\begin{equation}
d_3= - {1 \over 2 r r^\prime} \end{equation}
\begin{equation}
d_5=  {m_b^2 \over 4 r^3 r^\prime}+ {m_s^2 \over 4 r r^{\prime 3}} 
+ {1 \over 6 r^2 r^\prime}
+ {2m_s^2-m_s m_b-2q^2 \over 12 r^2 r^{\prime 2}} \;.
\label{d5_f1} \end{equation}

For $F_0(q^2)$ the formulae read as follows:
\begin{equation}
H=f_K f_B {M_B^2 \over m_b}(M_B^2-M_K^2) \label{h_f0} \end{equation}
\begin{equation}
\rho(s, s^\prime, q^2)
={3 \over 8 \sqrt{\lambda}} \left\{ \Delta(m_b -m_s) + {2 \Delta^\prime s 
-\Delta u \over \lambda} \left[m_b(2(\Delta-\Delta^\prime)+2s^\prime - u) + m_s(
u-2s)\right] \right\}
\label{rho_f0} 
\end{equation}
\begin{equation}
d_3=-{m_b (m_b- m_s) \over r r^\prime} \label{d3_f0} \end{equation}
\begin{eqnarray}
d_5&=&(m_b-m_s)\left [{m_b m_s^2 \over 2 r r^{\prime 3}}+
{m_b(m_b^2-m_b m_s+m_s^2-q^2) \over 6 r^2 r^{\prime 2}} \right.\nonumber \\
&+& \left. {m_b-m_s \over 6 r r^{\prime 3}}+
{m_b^3 \over 2 r^3 r^\prime}+{2 m_b \over 
3 r^2 r^\prime} \right ] \; . \label{d5_f0} \end{eqnarray}
In the formulae for the
coefficients of the non perturbative contributions, reported in this Appendix 
and in the following one, we have omitted all terms that vanish after 
the double Borel transform.

\newpage
\section{{\bf $B \to K^* \ell^+ \ell^-$} }

The quantities appearing in the sum rule for 
the form factor $T_1(q^2)$ in eq. (\ref{t1}) read as follows:
\begin{equation}
H=2 f_{K^*} M_{K^*} f_B {M_B^2 \over m_b} \label{h_t1}
\end{equation}

\begin{eqnarray}
\rho(s,s^\prime,q^2)&=& {3 \over 8 \sqrt{\lambda}} \Big\{ \Delta -{1 \over 
\lambda} [m_b m_s(2 \Delta^\prime s + 2 \Delta s^\prime
-u(\Delta+\Delta^\prime)) \nonumber \\
&+& (s^\prime-u)(2 \Delta^\prime s -\Delta u)
-s(2 \Delta s^\prime -\Delta^\prime u)-\Delta^{\prime 2} s 
-\Delta^2 s^\prime +\Delta \Delta^\prime u] \Big\}
\label{rho_t1} \end{eqnarray}
\begin{equation}
d_3=-{(m_b+m_s)\over 2 r r^\prime} \label{d3_t1} \end{equation}

\begin{eqnarray}
d_5&=&{m_s\over 12 r r^{\prime2}} +{ 3 m_b + 2 m_s \over 12 r^2 r^\prime} 
+{ m_s^2( m_b +  m_s) \over 4 r r^{\prime 3}} \nonumber \\
&+&{ m_b^2( m_b +  m_s) \over 4 r^3 r^\prime }
-{( m_b +  m_s)[m_b m_s +2(q^2-m_b^2-m_s^2)] \over 12 r^2 r^{\prime 2}} \; .
 \label{d5_t1} \end{eqnarray}
For the form factor $T_2(q^2)$:
\begin{equation}
H= - f_{K^*} M_{K^*} f_B {M_B^2 \over m_b}(M_B^2-M_{K^*}^2) \label{h_t2}
\end{equation}
\begin{equation}
\rho(s,s^\prime,q^2)= {3 \over 16 \sqrt{\lambda}} \left\{
m_b m_s (\Delta^\prime - \Delta) +(\Delta s^\prime-\Delta^\prime s) +
(u-2s) {(\Delta^{\prime 2} s+\Delta^2 s^\prime-\Delta \Delta^\prime u) \over \lambda} 
\right\}
\label{rho_t2} \end{equation}
\begin{equation}
d_3={(m_b - m_s) [(m_b+m_s)^2-q^2] \over 4 r r^\prime} \label{d3_t2}
\end{equation}
\begin{eqnarray}
d_5&=& -{3 m_b + m_s \over 24 r r^\prime}
-{(m_b - m_s)m_s^2[(m_b+m_s)^2-q^2]  \over 8 r r^{\prime 3}} 
-{(m_b - m_s)m_b^2[(m_b+m_s)^2-q^2]  \over 8 r^3 r^\prime }\nonumber \\
&+&{(m_b - m_s)[(m_b+m_s)^2-q^2][m_b m_s-2(m_b^2+m_s^2-q^2)]  
\over 24 r^2 r^{\prime 2}}\label{d5_t2} \\
&+& {q^2(2 m_b-m_s)-2(m_b+m_s)(m_b^2-m_s^2) \over 24 r r^{\prime 2}} +
 {q^2(5 m_b-4 m_s)-4(m_b+m_s)(m_b^2-m_s^2) \over 24 r^2 r^\prime } \; .
\nonumber \end{eqnarray}
Form factor $V(q^2)$:
\begin{equation}
H=f_{K^*} M_{K^*} f_B {M_B^2 \over m_b} {2 \over M_B + M_{K^*}} \label{H_V} 
\end{equation}
\begin{equation}
\rho=-{3 \over 4 \lambda^{3/2}}[m_b (2 \Delta s^\prime-\Delta^\prime u)+
m_s (2 \Delta^\prime s-\Delta u)] \label{rho_V} \end{equation}
\begin{equation}
d_3=-{1 \over r r^\prime} \label{d3_V} \end{equation}
\begin{equation}
d_5={1 \over 3 r^2 r^\prime} +{m_b^2 \over 2 r^3 r^\prime}
+{m_s^2 \over 2 r r^{\prime 3}}+{2(m_b^2+m_s^2-q^2)-m_b m_s \over 6 r^2 
r^{\prime 2}} \; .
\label{d5_V} \end{equation}
Form factor $A_1(q^2)$:
\begin{equation}
H=f_{K^*} M_{K^*} f_B {M_B^2 \over m_b} (M_B + M_{K^*}) \label{H_A1} 
\end{equation}
\begin{equation}
\rho={3 \over 8 \sqrt{\lambda}} \left[ (m_b \Delta^\prime +m_s \Delta)+ 
{2 m_b \over \lambda} (\Delta^{\prime 2} s +\Delta^2 s^\prime -\Delta 
\Delta^\prime u) \right] \label{rho_A1} 
\end{equation}
\begin{equation}
d_3=-{1 \over2 r r^\prime}[m_b^2+m_s^2-q^2+2m_b m_s] \label{d3_A1} 
\end{equation}
\begin{eqnarray}
d_5&=&-{1 \over 6 r r^\prime}+{3 m_b^2+9m_b m_s+4m_s^2-4q^2 \over 12 r^2 
r^\prime} +{4 m_b^2+6m_b m_s+6m_s^2-4q^2 \over 24 r^2 r^{\prime 2}} \nonumber 
\\
&+& {m_s^2 [(m_b+m_s)^2-q^2] \over 4 r r^{\prime 3}}+
{m_b^2 [(m_b+m_s)^2-q^2] \over 4 r^3 r^\prime} \nonumber \\
&-& {-4 m_b^4-6m_b^3 m_s-4 m_b^2 m_s^2-6m_b m_s^3-4 m_s^4-4m_b^2 q^2+6 m_b m_s 
q^2+8 m_s^2 q^2 -4 q^4 \over 24 r^2 r^{\prime 2}} \; . \label{d5_A1} 
\end{eqnarray}
Form factor $A_2(q^2)$:
\begin{equation}
H=f_{K^*} M_{K^*} f_B {M_B^2 \over m_b} {1 \over M_B + M_{K^*}} \label{H_A2} 
\end{equation}
\begin{eqnarray}
\rho&=&-{3 \over 8 \lambda^{3/2}} \left\{m_b(2 \Delta s^\prime -\Delta^\prime u)
+m_s(2 \Delta^\prime s -\Delta u) \right.\nonumber \\
&+&\left. {2 m_b \over \lambda}\right [\Delta^{\prime 2}(2 s s^\prime +u^2-3 s u)+3\Delta^2(2 
s^{\prime 2}-s^\prime u)+2 \Delta \Delta^\prime (-3 s^\prime u+2 s s^\prime+u^2
)\left ] \right\} \label{rho_A2} \end{eqnarray}
\begin{equation}
d_3=-{1 \over2 r r^\prime} \label{d3_A2} 
\end{equation}
\begin{equation}
d_5=-{1 \over 6 r^2 r^\prime}+{m_s^2 \over 4 r r^{\prime 3}}+
{m_b^2 \over 4 r^3 r^\prime}+{2 m_b^2+2 m_s^2- 2 q^2-m_b m_s \over 12 r^2 
r^{\prime 2}} \; . \label{d5_A2} \end{equation}

\clearpage

\begin{table}
\caption{Wilson coefficients $C_i(\mu)$ for 
$\Lambda^{(5)}_{\overline {MS}}=225 \; MeV$, $\mu =5 \; GeV$ and $m_t=174 
\; GeV$.
\\}
\begin{tabular}{ccc}
 & NDR & HV  \\ \tableline
$C_1$ & $-0.243$ & \\
$C_2$ & $1.105$ & \\
$C_3$ & $1.083 \times 10^{-2}$ & \\
$C_4$ & $-2.514 \times 10^{-2}$ & \\
$C_5$ & $7.266 \times 10^{-3}$ & \\
$C_6$ & $-3.063 \times 10^{-2}$ & \\
$C_7$ & $-0.312$ & \\
$C_9$ & $4.193 $ & $3.998$ \\
$C_{10}$ & $-4.578$ & \\
\end{tabular}

\end{table}

\vskip 2cm

\begin{table}
\caption{Parameters of the form factors. 
The functional $q^2$ dependence is either polar:
$F(q^2)=F(0)/(1-q^2/M_P^2)$ or linear:  $F(q^2)=F(0)(1+ \beta q^2)$.
For the form factor $F_T$ see text.
\\}
\begin{tabular}{cccc}
 & $F(0)$ & $M_P \; (GeV)$ & $\beta \; (GeV^{-2})$  \\ \tableline
$F_1$ & $0.25 \pm 0.03$ & $5$ &\\
$F_0$ & $0.25 \pm 0.03$ & $7$ &\\
$F_T$ & $-0.14 \pm 0.03$ & &\\
$V$ & $0.47 \pm 0.03$ & $5$ &\\
$A_1$ & $0.37 \pm 0.03$ & & $-0.023$ \\
$A_2$ & $0.40 \pm 0.03$ & & $0.034$ \\
$A_0$ & $0.30 \pm 0.03$ &$4.8$ & \\
$T_1$ & $0.19 \pm 0.03$ &$5.3$ & \\
$T_2$ & $0.19 \pm 0.03$ & & $-0.02$ \\
$T_3$ & $-0.7$ & & $0.005$ \\
\end{tabular}

\end{table}

\clearpage

\hskip 3 cm {\bf FIGURE CAPTIONS}
\vskip 1 cm
{\bf Fig. 1} \par
\noindent 
Form factors 
$F_1(q^2)$, $F_0(q^2)$ and $F_T(q^2)$ of the transition 
$B \to K \ell^+ \ell^-$.
The curves refer to different sets of parameters:
$s_0=33 \; GeV^2$ and $s^\prime_0=1.4 \; GeV^2$ (continuous line),
$s_0=33 \; GeV^2$ and $s^\prime_0=1.6 \; GeV^2$ (dashed line),
$s_0=36 \; GeV^2$ and $s^\prime_0=1.4 \; GeV^2$ (dotted line),
$s_0=36 \; GeV^2$ and $s^\prime_0=1.6 \; GeV^2$ (dashed-dotted line).
The Borel parameters are fixed to 
$M^2=8 \; GeV^2$, $M^{\prime 2}=2  \; GeV^2$.
\vskip 1 cm
{\bf Fig. 2} \par
\noindent 
Form factors 
$T_1(q^2)$, $T_2(q^2)$ and $T_3(q^2)$ of the transition 
$B \to K^* \ell^+ \ell^-$.
The curves refer to different sets of parameters:
$s_0=33 \; GeV^2$ and $s^\prime_0=1.6 \; GeV^2$ (continuous line),
$s_0=33 \; GeV^2$ and $s^\prime_0=1.8 \; GeV^2$ (dashed line),
$s_0=36 \; GeV^2$ and $s^\prime_0=1.6 \; GeV^2$ (dotted line),
$s_0=36 \; GeV^2$ and $s^\prime_0=1.8 \; GeV^2$ (dashed-dotted line).
The Borel parameters are fixed to 
$M^2=8 \; GeV^2$, $M^{\prime 2}=2  \; GeV^2$.
\vskip 1 cm
{\bf Fig. 3} \par
\noindent 
Form factors $V(q^2)$,  $A_1(q^2)$, $A_2(q^2)$ and $A_0(q^2)$ of
$B \to K^* \ell^+ \ell^-$.
The curves refer to the same set of parameters as in fig.2.
\vskip 1 cm
{\bf Fig. 4} \par
\noindent 
Momentum dependence of the ratio between
rare and semileptonic form factors
${\cal R}=F_i(q^2)/F_i^{IW}(q^2)$; $F_i^{IW}$ are obtained from 
eqs.(\ref{is_ft}-\ref{is_t2}).  
\vskip 1 cm
{\bf Fig. 5} \par
\noindent 
Invariant mass squared distribution of the lepton pair for the decay 
$B \to K \ell^+ \ell^-$: the dashed line refers to the short distance 
contribution only.
\vskip 1 cm
{\bf Fig. 6} \par
\noindent 
Forward-backward asymmetry in the decay $B \to K^* \ell^+ \ell^-$; the dashed 
line refers to the short distance contribution only.
\vskip 1 cm
{\bf Fig. 7} \par
\noindent 
Invariant mass squared distribution of the lepton pair for the decay 
$B \to K^* \ell^+ \ell^-$: the dashed line refers to the short distance 
contribution only.
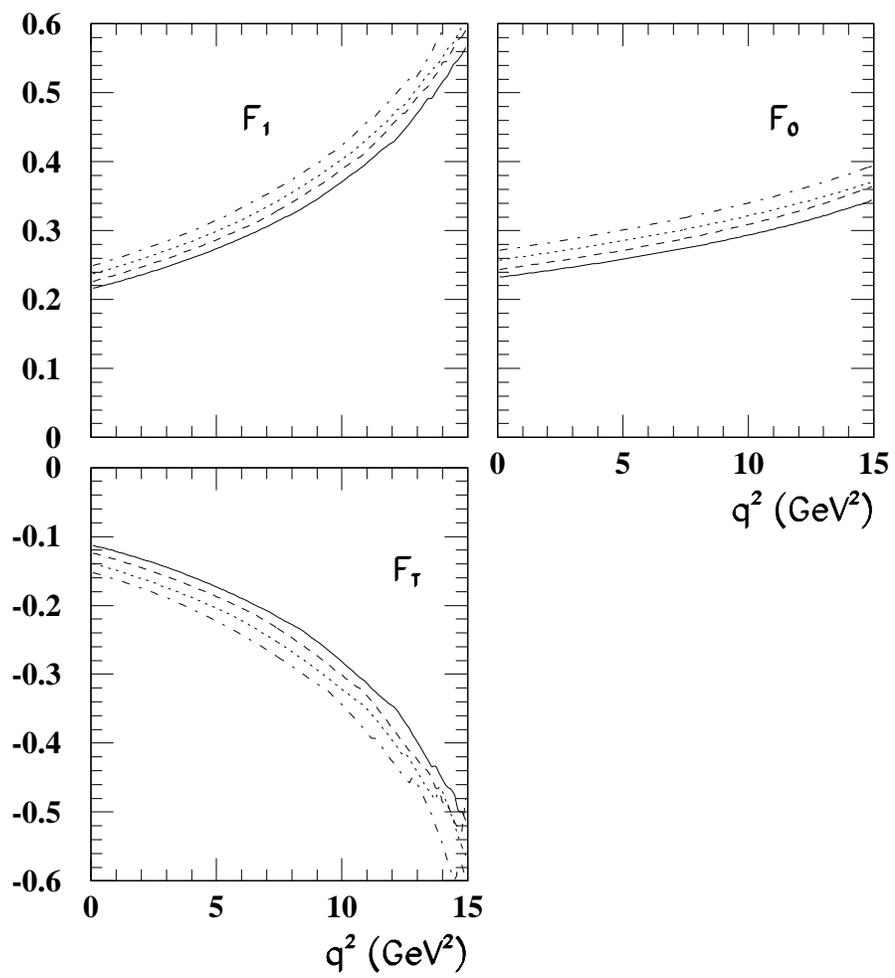
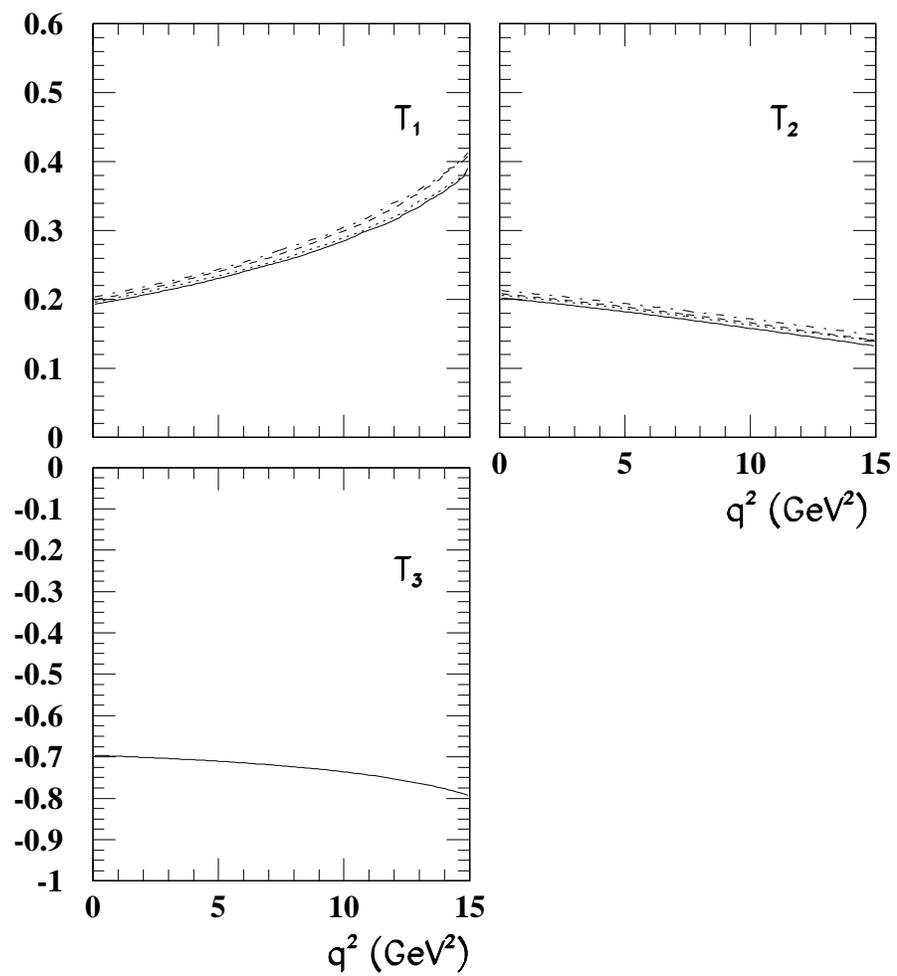
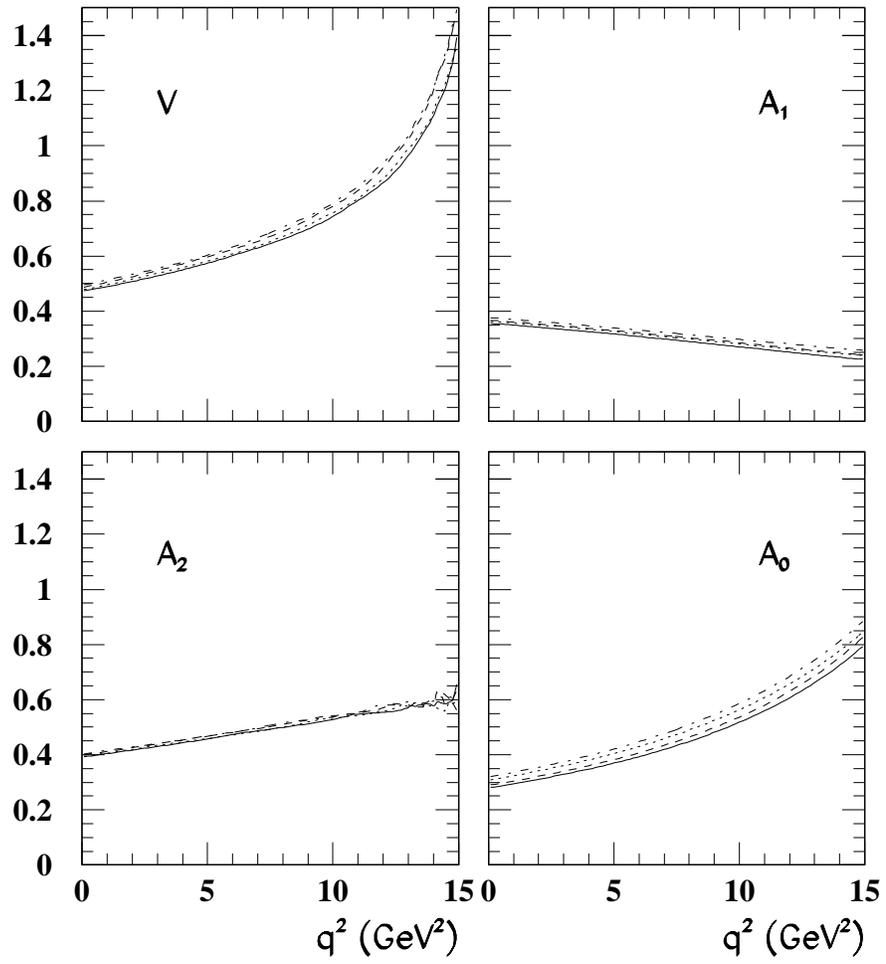
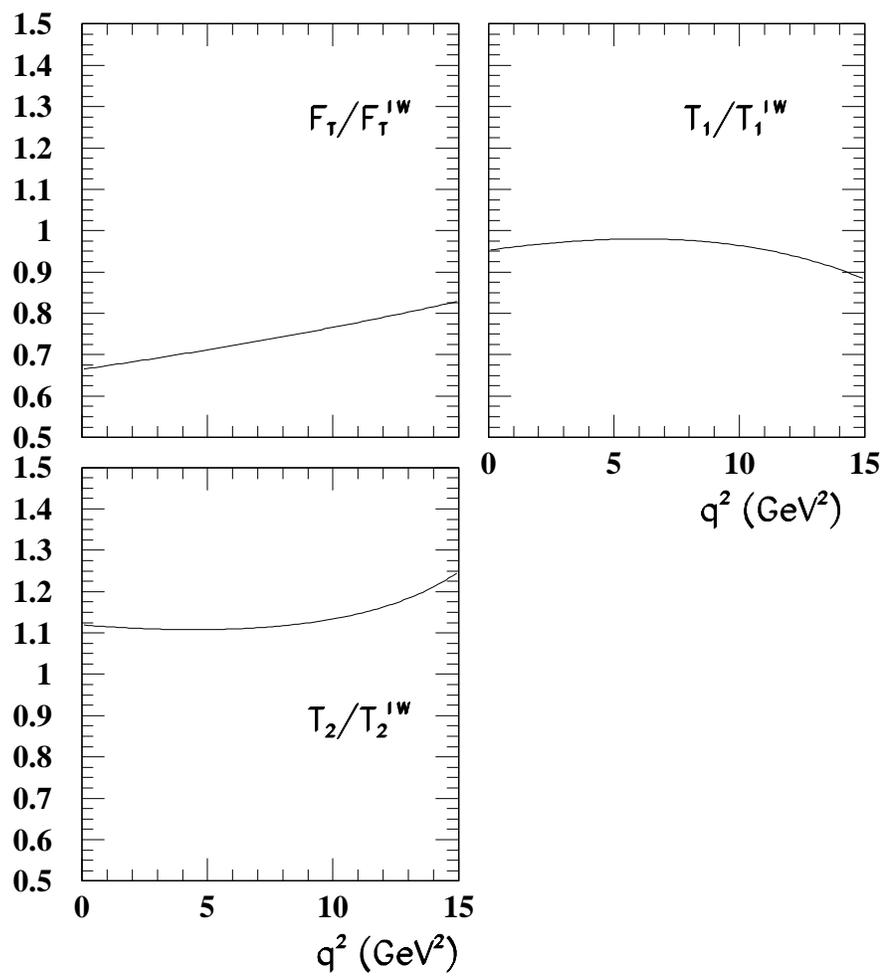
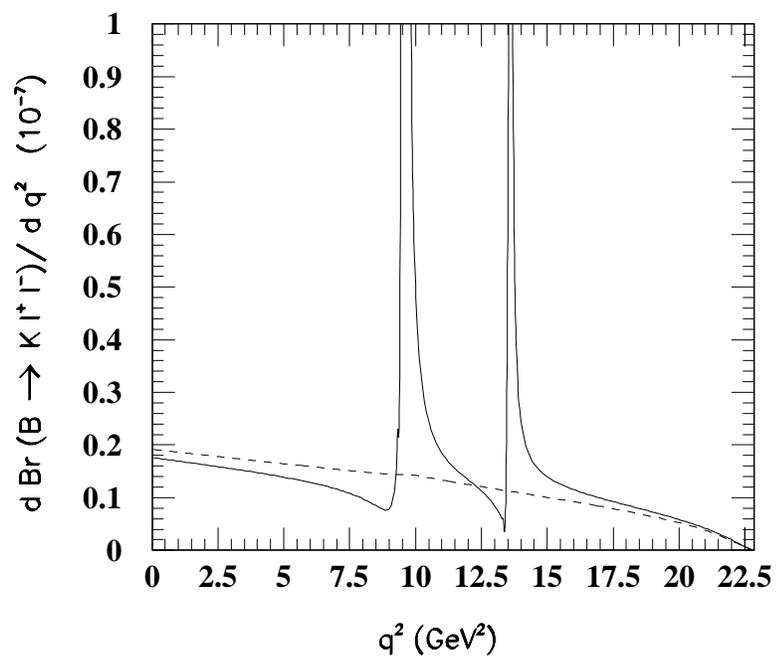
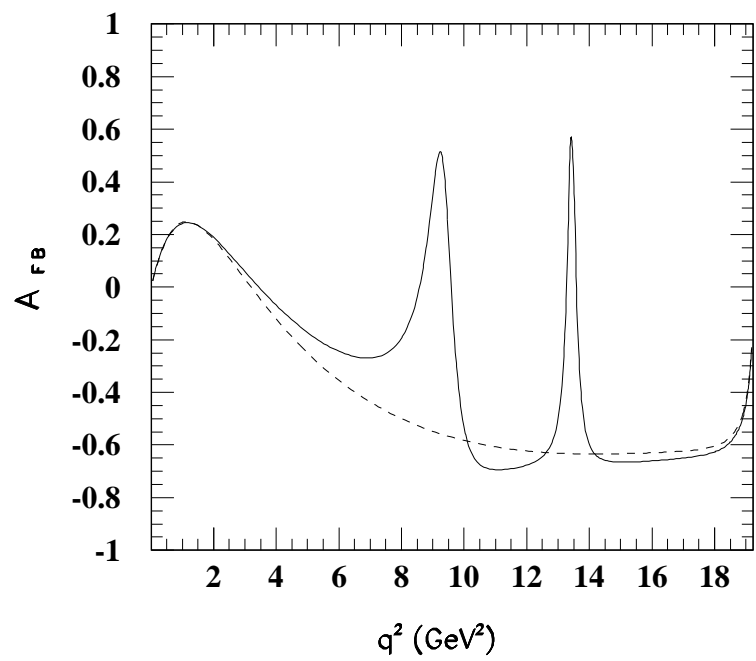
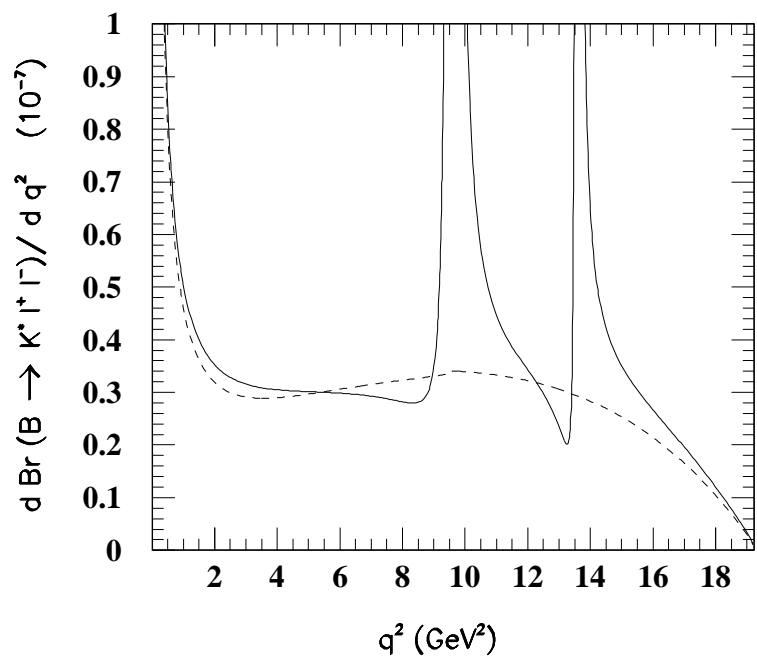
\end{document}